\documentclass[aps,prl,twocolumn,superscriptaddress,showpacs,reprint]{revtex4-1}
\usepackage{graphicx}
\usepackage{mathrsfs}
\usepackage{amsmath}
\usepackage{subfigure}
\usepackage{bm}
\usepackage{verbatim}
\usepackage{color}
\usepackage{xcolor}
\usepackage{hyperref}
\usepackage{epsfig}
\usepackage{longtable}

\begin{document}
\title{Quantum Anomalous Hall Insulator of Composite Fermions}
\author{Yinhan Zhang}
\affiliation{Institute of Physics, Chinese Academy of Sciences, Beijing 100190, China}
\author{Junren Shi}
\affiliation{International Center for Quantum Materials, Peking University, Beijing 100871, China}
\affiliation{Collaborative Innovation Center of Quantum Matter, Beijing 100871, China}

\date{\today}

\begin{abstract}
We show that a weak hexagonal periodic potential could transform a two-dimensional electron gas with an even-denominator magnetic filling factor to a quantum anomalous Hall insulator of composite fermions, giving rise to fractionally quantized Hall effect. The system provides a realization of the Haldane honeycomb-net model, albeit in a composite fermion system. We further propose a trial wave function for the state, and numerically evaluate its relative stability against the competing Hofstadter state. Possible sets of experimental parameters are proposed. 
\end{abstract}
\maketitle

It was long predicted that a magnetic insulator could exhibit a nonzero quantized Hall conductance in the absence of an external magnetic field, known as the quantum anomalous Hall insulator (QAHI)~\cite{Haldane-model}.  Searching for QAHI in real or artificial materials has been the focus of intensive recent theoretical investigations~\cite{Onoda2003,Liu2008,qiao2010,Yu2010,qiao2012,zhang2012}. Experimentally, it is demonstrated recently that a Cr-doped (Bi,Sb)$_2$Te$_3$ thin film could be the first QAHI ever realized~\cite{chang2013}.   As a next logical development of theory, the possibility of finding a QAHI exhibiting the fractional quantum Hall effect in the presence of strong electron-electron interaction is extensively discussed in the context of the flat Chern band insulator~\cite{Tang2011,Sun2011,Neupert2011,Sheng2011,Bergholtz2013}, which is yet to be materialized. 
  
In this Letter, we propose a new class of QAHI that exhibits the fractional quantum Hall effect, i.e., quantum anomalous Hall insulator of composite fermions (CF-QAHI). The composite fermion (CF), which is defined as an electron binding with two or even number of quantized vortices, is not only the key theoretical apparatus for describing the extremely complex fractional quantum Hall states, but also a physical entity with well defined charge, spin, and statistics~\cite{Jains,Heinonen,Wen2004}.  In particular, a two-dimensional electron gas (2DEG) with an even-denominator magnetic filling factor behaves just like a fermi liquid of CFs at the zero effective magnetic field, with the external magnetic field fully compensated by the quantized vortices bound in CFs~\cite{HLR,Kalmeyer1992,Willett1993,Kang1993a,Goldman1994,Smet1996}. Moreover, it was experimentally demonstrated that the picture of CFs is valid even in the presence of superstructures such as anti-dot arrays~\cite{Kang1993a} and periodic potentials~\cite{Smet1999,Oppen1998,Mirlin1998}. An interesting possibility naturally arises: a carefully designed superstructure could transform the CF fermi liquid to a CF band insulator with nontrivial topology~\cite{TKNN}. We call the resulting insulator as a CF-QAHI.

We show that this is indeed possible. We investigate the effect of a weak hexagonal periodic potential superimposed on a spinless 2DEG system at an even-denominator magnetic filling factor $\nu_{M}={1}/{2p}$.  By employing the CF-mean field theory, we show that band gaps can be opened when the strength of the periodic potential exceeds a critical value.  More importantly, a staggered effective magnetic field ${B}^{\ast}$ experienced by CFs emerges from the locally incomplete compensation of the external magnetic field,  making the system a natural realization of the Haldane honeycomb-net model~\cite{Haldane-model}, albeit for CFs.  When the spatial period  of the potential commensurates with the magnetic length and each unit cell contains an integer number of electrons, the system becomes a CF-QAHI, with a Hall conductance fractionally quantized at $-1/(2p-1)(e^2/h)$. We further propose a CF many-body trial wave function, and show that CF ground state can substantially lower the  Coulomb interaction energy, and could be stabilized in strong-interacting limit. Based on these calculations, realistic sets of experimental parameters could be proposed.  

\begin{figure}[tb]
\centering
\includegraphics[width=0.60\columnwidth]{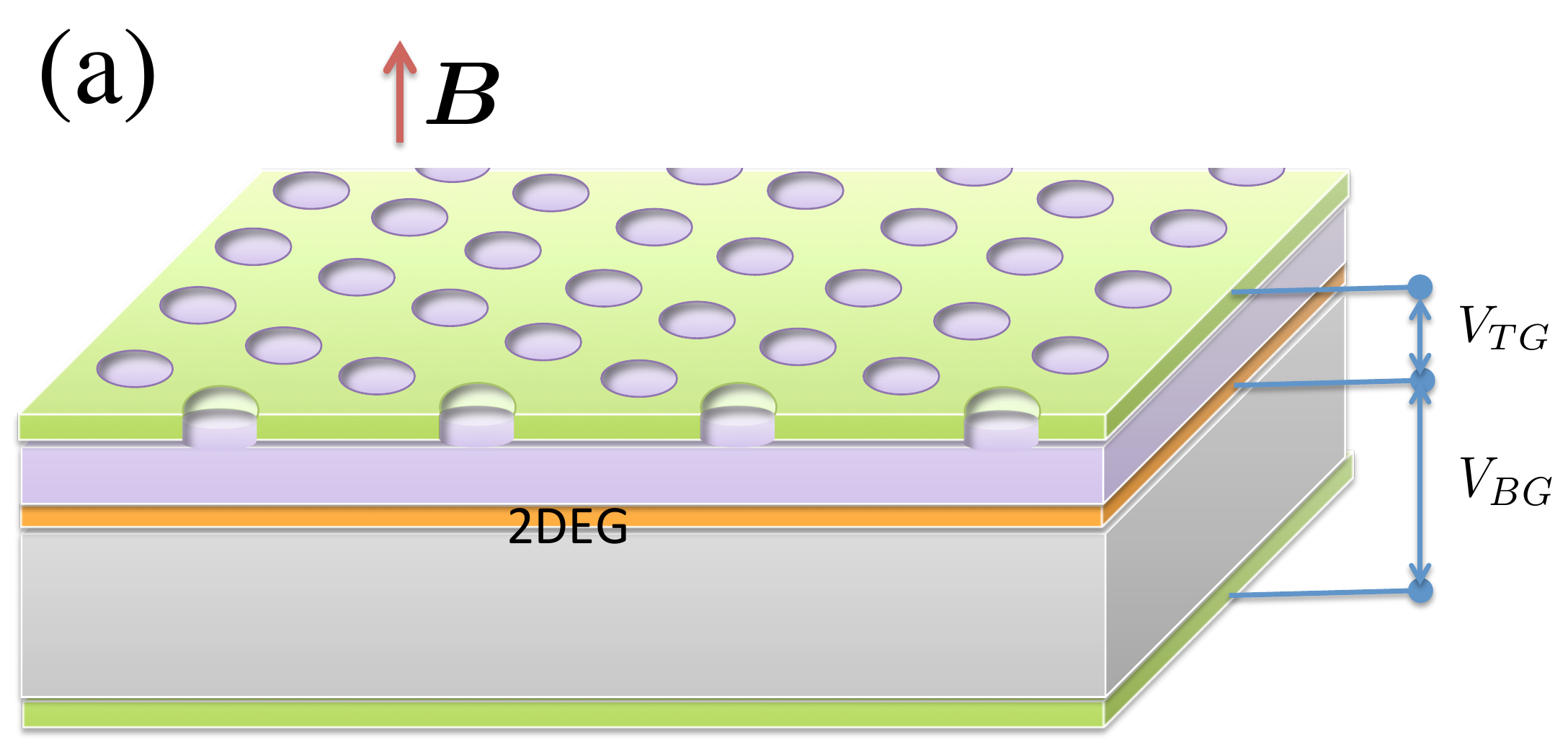}
\includegraphics[width=0.38\columnwidth]{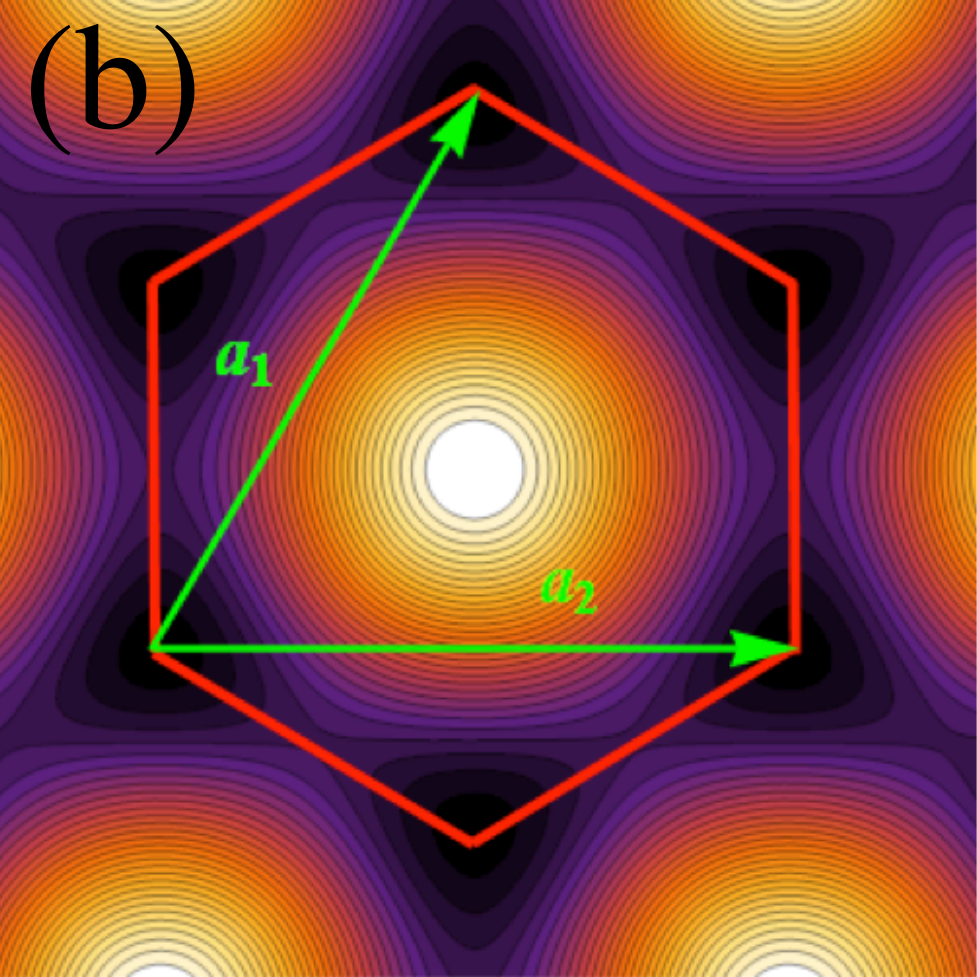}
\caption{(color online) (a) The schematic structure of the proposed device: a 2DEG confined in a quantum well or heterostructure, with both a back gate and a top gate. The top gap is patterned with anti-dot array of triangular lattice. The back and top gate voltages $V_{BG}$ and $V_{TG}$ can be independently tuned to control the electron density and modulation strength of the periodic potential $V_{0}$. A magnetic field $\bm{B}$ is applied perpendicularly to the 2DEG. (b) The spatial profile of the periodic potential.}
\label{fig:illustration}
\end{figure}

We consider a system shown in Fig.~\ref{fig:illustration}(a): a 2DEG system with both back gate and top gate. The top gate is patterned with an anti-dot array of triangular lattice, and when the top gate voltage is applied, superimposes a weak hexagonal periodic potential on the 2DEG. The potential profile can be modeled as,
\begin{multline}
U(\bm{r})  =-\frac{1}{2}V_{0} \left\{\cos\left( (\bm{b}_{1}+\bm{b}_{2})\cdot\bm{r}-\frac{\pi}{3}\right) \right. \\
\left. +\cos\left(\bm{b}_{1}\cdot\bm{r}+\frac{\pi}{3}\right) +\cos\left(\bm{b}_{2}\cdot\bm{r}+\frac{\pi}{3}\right) \right\} +\frac{3V_{0}}{2},
\label{potential}  
\end{multline}
where $\bm{b}_{1}=({4\pi}/{3a})(0,1)$ and
$\bm{b}_{2}=({2\pi}/{3a})(\sqrt{3},-1)$ are the reciprocal lattice vectors, with $a$ being lattice constant of hexagonal structure, and $V_{0}$ the strength of modulation of the periodic potential. The potential profile is shown in Fig.~\ref{fig:illustration}(b). It has two valleys in each unit cell, mimicking the structure of graphene.  By tuning the back and top gate voltages, the carrier density $n_e$ and modulation strength of the periodic potential $V_0$ can be controlled independently. The 2DEG system is subjected to a perpendicular magnetic field $\bm B$. We assume that the magnetic field is strong enough to fully lift the spin degeneracy of electrons. As a result, electrons are considered as spinless. In the following discussions, the system is always kept at a magnetic filling factor $\nu_{M}={1}/{2p}$ with $p$ being an integer, i.e.,  the magnetic field strength is always kept proportional to the electron density:  $B_z=2p n_e \phi_0$, where $\phi_0=hc/|e|$ is the quantum of magnetic flux. 

\begin{figure}[tb]
\centering
\includegraphics[width=0.51\columnwidth]{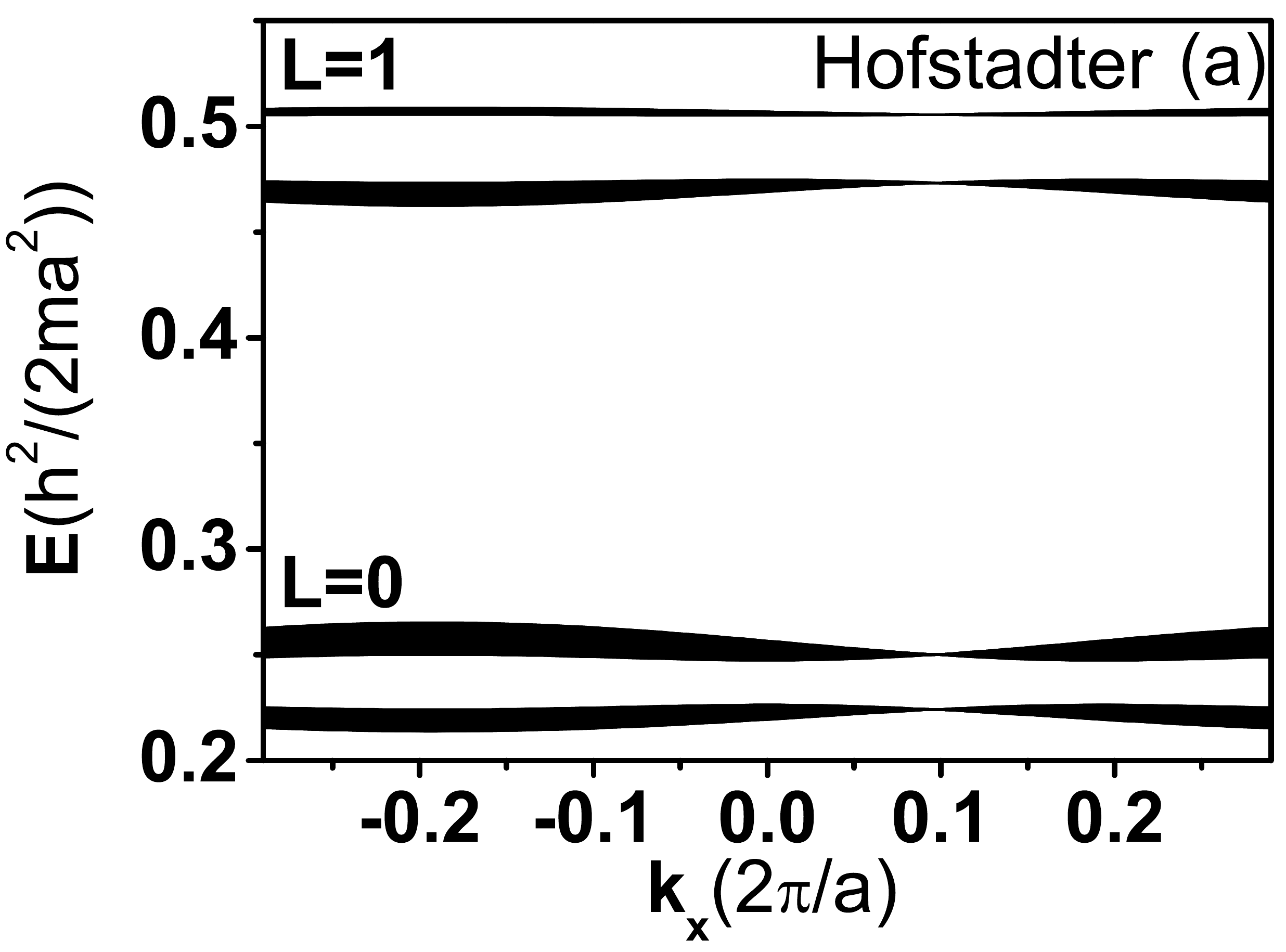}
\includegraphics[width=0.46\columnwidth]{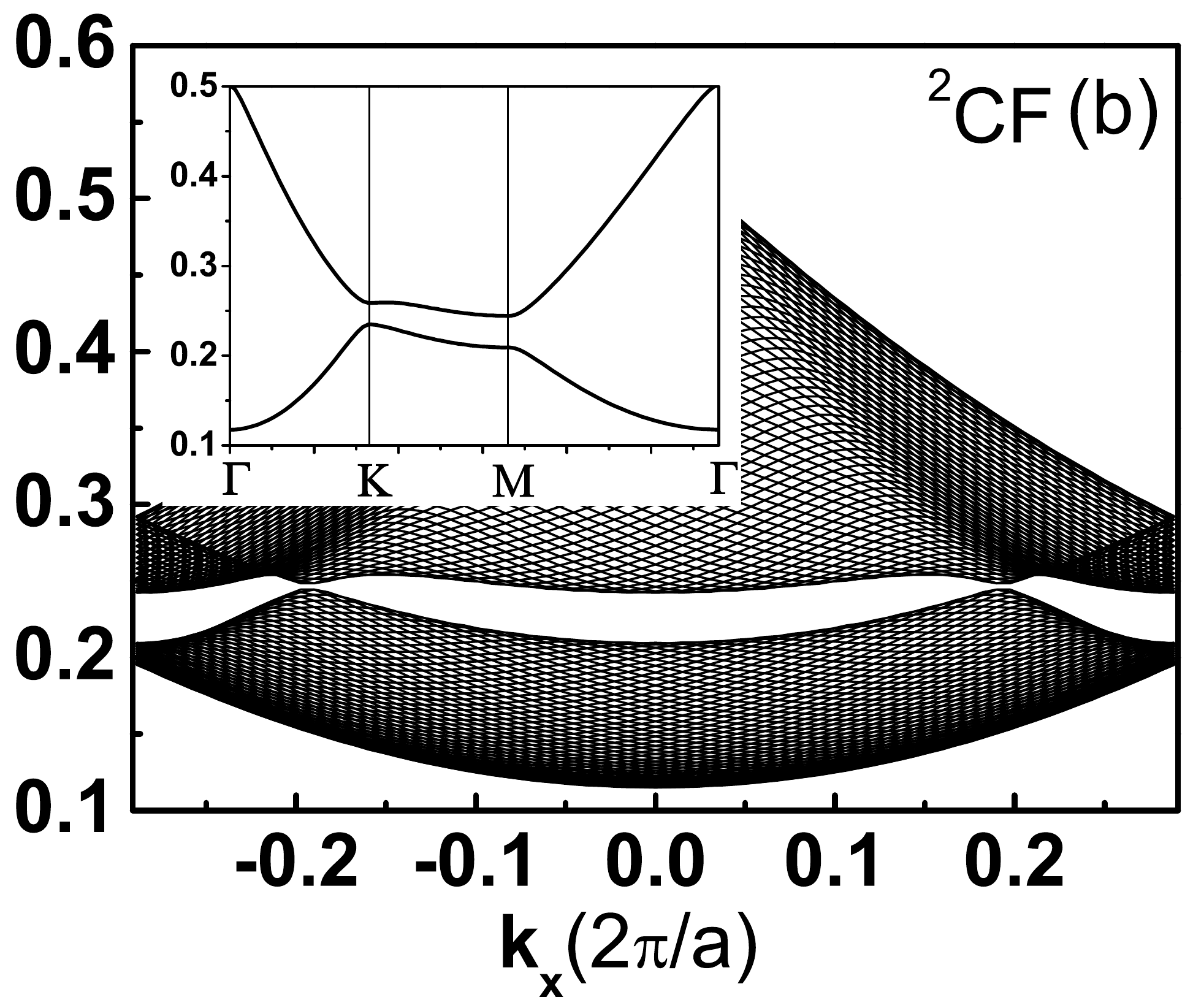}
\caption{(color online) (a) Electron spectrum at magnetic filling factor $1/2$ in the presence of the hexagonal periodic potential with $V_{0}=0.08\epsilon_{a}$ ($\epsilon_{a}\equiv h^{2}/(2m_{b}a^{2})$) and $\nu_{e}=1$, projected to $k_{x}$-axis. $L=0(1)$ denotes the index of the original Landau level, which is split to sub-bands in the presence of the periodic potential. (b) Projected CF spectrum at the same parameters. Inset: CF $\Lambda$-level dispersion along $\Gamma$-$K$-$M$-$\Gamma$ direction in the hexagonal Brillouin zone. The spectra are calculated using the periodic boundary condition.}
\label{fig:MEB}
\end{figure}

We calculate the Hofstadter spectra of the system in the absence of electron-electron interaction~\cite{Hofstadter1976,MacDonald1984}. We assume that each unit cell has one electron, i.e., the band filling factor $\nu_e = 1$. The lattice constant is chosen to be $a=(2/\sqrt{3})^{3/2}\sqrt{\pi p \nu_e} l_{m}$ to have $\nu_{M}={1}/{2p}$, where $l_{m}=\sqrt{\hbar c/|e|B_{z}}$ is the magnetic length. 
The band structure is shown in Fig.~\ref{fig:MEB}(a). In the presence of the periodic potential, the lowest Landau level is spread to a set of sub-bands ($L=0$). When the potential is weak, the total spreading width of energy $w_{L}$ is much smaller than the Landau level spacing $\hbar\omega_{0}$, as shown in Table.~\ref{table:scales}. 

In the presence of electron-electron interaction, a new energy scale for Coulomb interaction $E_{C}\equiv e^{2}/(\epsilon l_{m})$ emerges, where $\epsilon$ is the relative dielectric constant of the 2DEG. When the interaction energy dominates over the band spreading $w_{L}$, a collection of normal electrons is not a good representation of the true ground state of the system anymore. The system will be in the regime of the fractional quantum Hall, and the proper description of the system should be a collection of CFs. The ratio $E_{C}/w_{L}$ are listed in Table.~\ref{table:scales}. One can see that the interaction energy always dominates over the band spreading in electron systems with $r_{s}\gg 1$, and becomes more prominent in the smaller filling factors (larger $p$). It is thus mandatory to discuss the problem in the CF-picture.
 
\begin{table}[tbp]
\caption{The comparison of different energy scales of cyclotron energy $\hbar\omega_0$, energy spreading width of the lowest Landau level $w_L$ and the interaction energy scale $E_c$ at the critical value of $V_{0}$ to insulating phases $V_{c}$ (see Fig.~\ref{fig:PD}) for $\nu_{e}=1$ and $\nu_{M}=1/2p$. The strength of the interaction is characterized by a dimensionless density parameter $r_{s}\equiv 1/(a^\ast_{B}\sqrt{\pi n_{e}})$ with $a^{\ast}_{B}=\epsilon \hbar^{2}/(m_{b}e^{2})$ being the effective Bohr radius of 2DEG.}
\begin{ruledtabular}
\begin{tabular}{|c|c|c|c|}
\hline
 $p$ & $V_{c}/\hbar\omega_{0}$ &  $w_{L}/\hbar\omega_{0}$ & $E_{C}/w_{L}$\tabularnewline
 \hline

 $p=1$ & $0.23$ &  $0.158$ & $3.16r_{s}$\tabularnewline
 \hline
 $p=2$ & $0.078$ & $0.081$ & $ 4.37r_{s}$\tabularnewline
 \hline
 $p=3$ & $0.033$ & $0.054$ & $5.35r_{s} $\tabularnewline
 \hline
 $p=4$ & $0.0148$ & $0.022$ &$11.4r_{s} $  \tabularnewline
 \hline
 $p=5$ & $0.0069$ & $0.011$ & $20.4r_{s} $\tabularnewline
 \hline
 \end{tabular}
 \end{ruledtabular}
 \label{table:scales}
 \end{table}

To catch the essence of CF physics, we employ the CF-mean field theory, which describes the system in an effective Chern-Simon CF Hamiltonian~\cite{HLR},
\begin{align}
\hat{H}_{CSCF}=\frac{1}{2m_{b}}(\bm{\hat{p}}-\frac{e}{c}\bm{A}^{\ast}(\bm{r}))^{2}+U^{\ast}(\bm{r}),
\end{align}
where, $\bm{A}^{\ast}(\bm{r})$ is the residual vector potential, giving rise to an effective magnetic field experienced by CFs,
\begin{align}
\bm{B}^{\ast}(\bm{r})\equiv\nabla\times\bm{A}^{\ast}(\bm{r})=\bm{B}-2p\phi_{0}\rho(\bm{r})\hat{z},
\label{Effectvector}
\end{align}
which has two parts of contribution: one is from the external magnetic field $\bm{B}$, and the other from the quantized vortices bound by CFs. Although the external magnetic field is spatially uniform, the effective magnetic field is inhomogeneous due to the modulation of electron density $\rho(\bm{r})$. It defines an effective staggered magnetic field that has the same periodicity as the hexagonal potential and the zero total effective magnetic flux through each unit cell. The effective scalar potential $U^{\ast}(\bm{r})$ also has two contributions: one is from the external potential, and the other is from the Hartree-like self-consistent Coulomb interaction potential (screening effect). To simplify our calculation, we assume that the $U^{\ast}(\bm{r})$ has the same form as $U(\bm{r})$ defined in Eq.~(\ref{potential}).  In our calculation, we solve CF eigen-problem in the plane wave basis with $25$ reciprocal lattice vectors, and the effective staggered magnetic field $\bm{B}^{\ast}(\bm{r})$ is self-consistently determined by Eq.~(\ref{Effectvector}) iteratively.

Figure~\ref{fig:MEB}(b) shows an example of the CF spectrum. One can see that an indirect gap develops between $\bm{K}$-point and $\bm M$-point of the Brillouin zone. The system becomes a CF-insulator.  We calculate the Chern number $\mathcal{C}$ contributed by the band below the gap~\cite{Xiao2010},  and find that $\mathcal{C}=-1$. As a result, the system is a CF-QAHI.

\begin{figure}[tb]
\centering
\includegraphics[width=0.43\columnwidth]{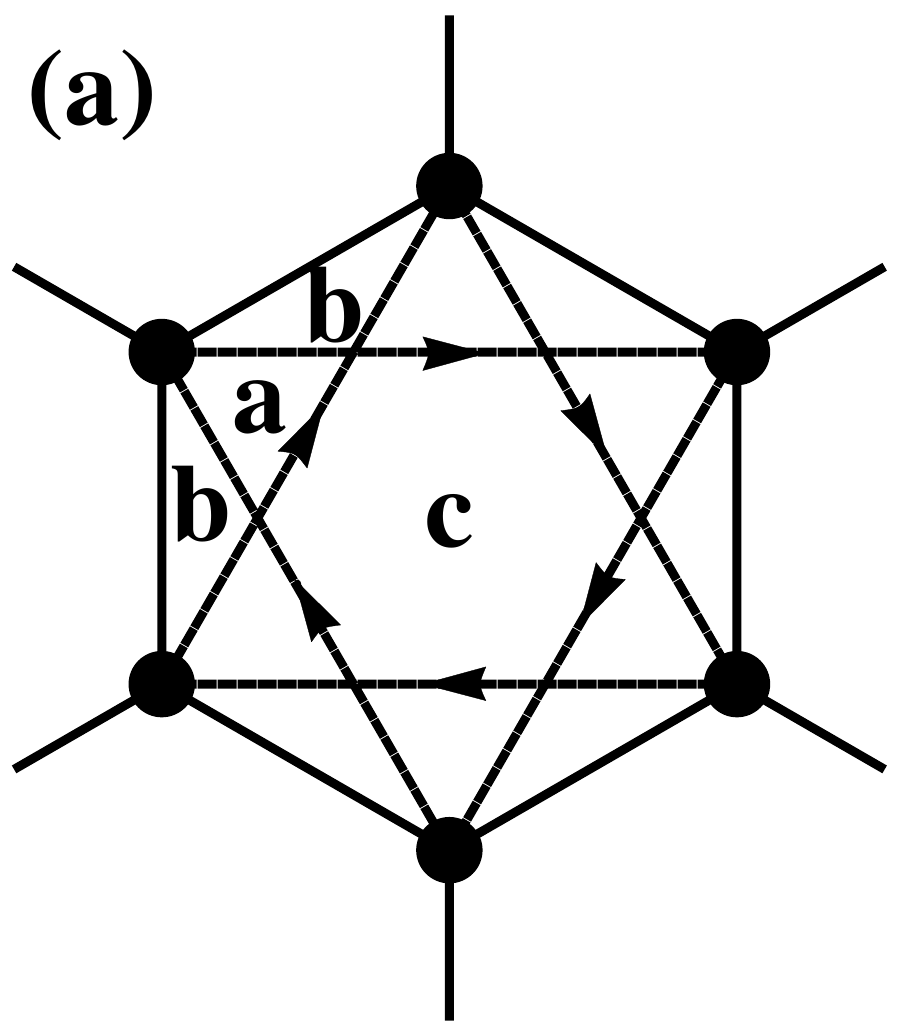}
\includegraphics[width=0.55\columnwidth]{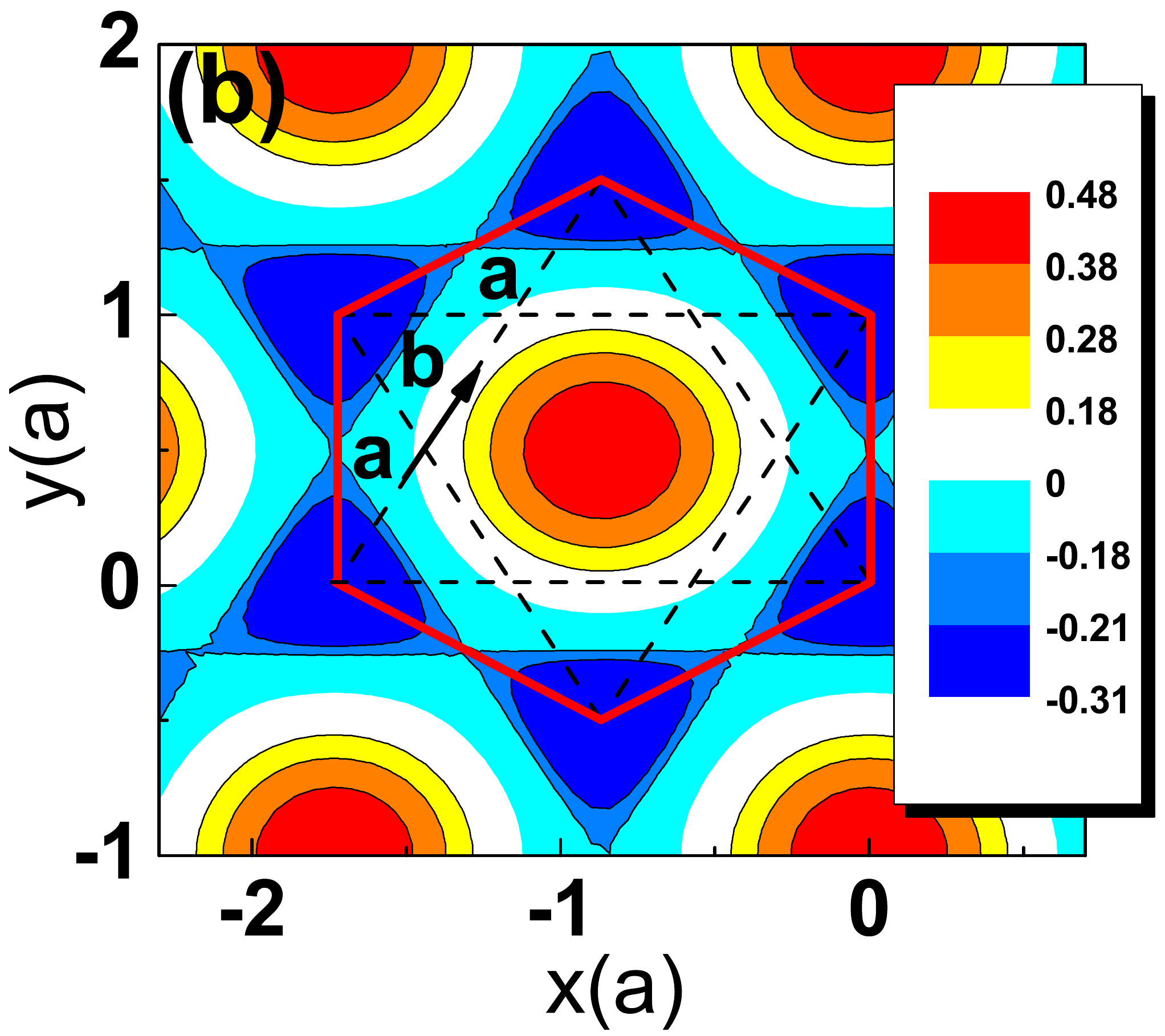}
\caption{(color online) (a) The Haldane honeycomb-net model with the next-nearest hopping and staggered magnetic flux. The next-nearest hopping constant gains a phase factor $\phi=-2\pi(2\Phi_{a}+\Phi_{b})/\phi_{0}$, where $\Phi_{a(b)}$ is the magnetic flux through the plaquette $a$ ($b$).  (b) The distribution of the staggered magnetic field of a CF insulating phase at $V_{0}=0.08\epsilon_a$, $p=1$, and $\nu_{e}=1$. }
\label{fig:Haldanemodel}
\end{figure}

The underlying physics can be understood by a mapping from our system in the continuous space to the Haldane honeycomb-net model (see Fig.~\ref{fig:Haldanemodel}(a)), which is the first proposed model for QAHI~\cite{Haldane-model}. The hexagonal potential has two valleys in each unit cell, defining sub-lattice sites of honeycomb network. The effective staggered magnetic field $\bm{B}^{\ast}(\bm{r})$ provides the required periodic local magnetic flux density with the zero total flux through the unit cell. Figure~\ref{fig:Haldanemodel}(b) shows that the spatial distribution of $\bm{B}^{\ast}(\bm{r})$. We can determinate the phase parameter $\phi=0.15\pi$. From Ref.~\onlinecite{Haldane-model}, the system should have a quantized Hall conductance ${e^{2}}/h$, corresponding to $\mathcal{C}=-1$, consistent with our result.  Our system thus provides a natural realization of the Haldane honeycomb-net model, albeit for CFs.

We find that the system can be transformed to a CF-QAHI only when the value of $V_{0}$ exceeds a critical value $V_{c}$. Although an infinitesimal  $V_{0}$ is sufficient to open a gap at $K$-point, the band dispersion at other quasi-momentum prevents the opening of a full gap. This can already be seen in the inset of Fig.~\ref{fig:MEB}, where the gap at $K$-point is overshadowed by the conduction band near the $M$-point, resulting in an indirect band gap.  

\begin{figure}[t]
\centering
\includegraphics[width=0.49\columnwidth]{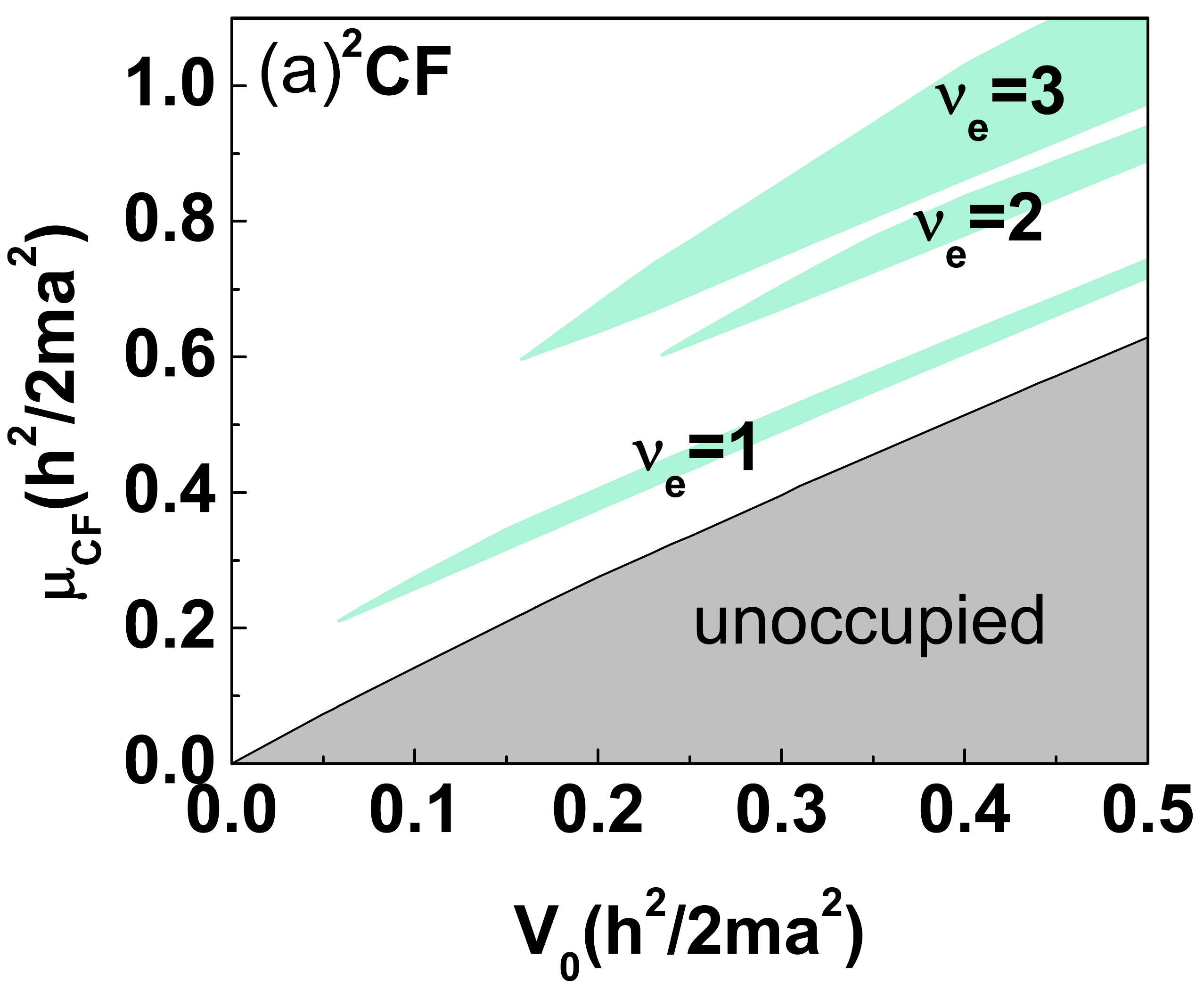}
\includegraphics[width=0.49\columnwidth]{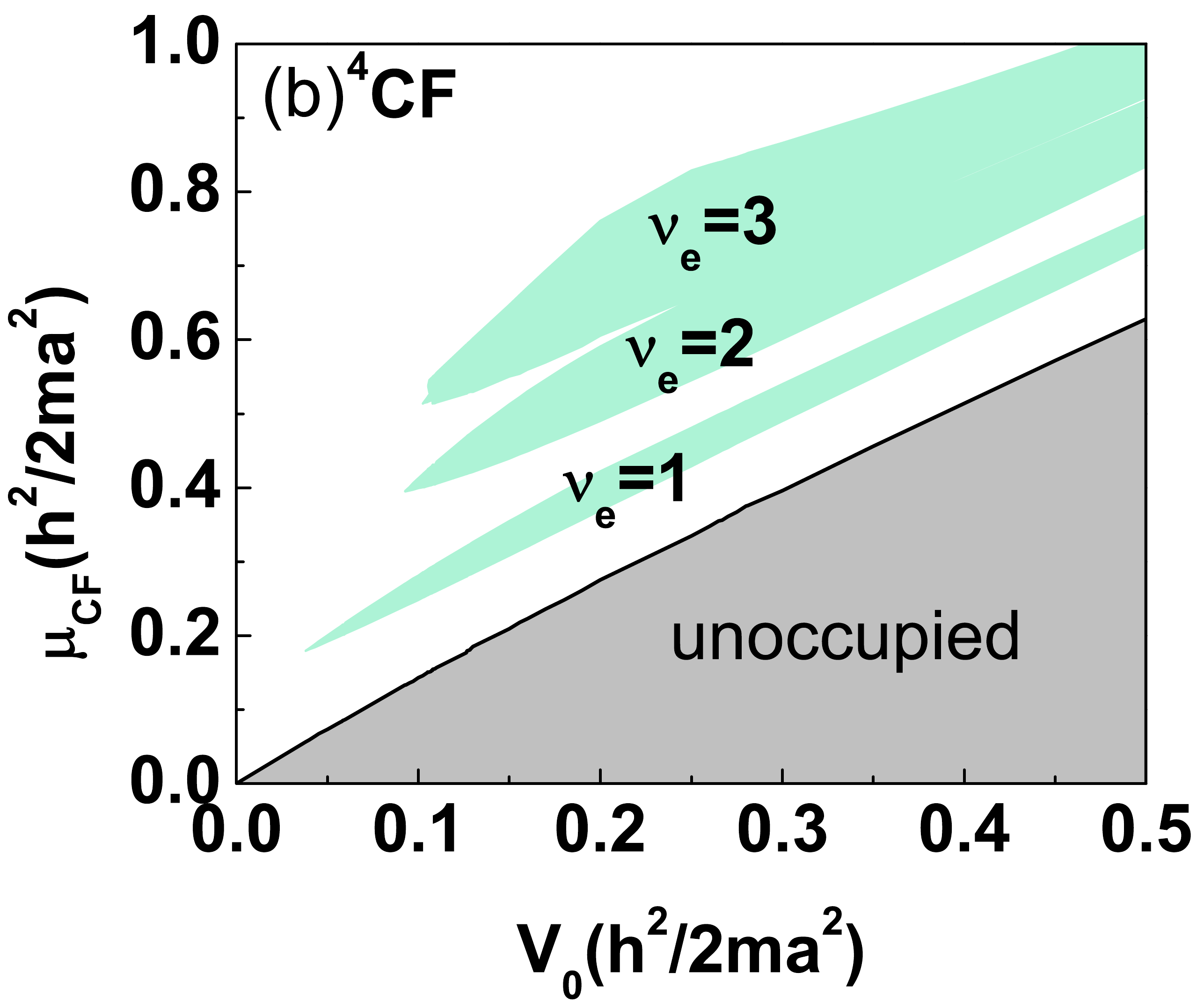}
\caption{(color online) CF metal-insulator phase diagrams in $V_{0}$-$\mu_{CF}$ for (a) $p=1$  and (b) $p=2$. The cyan zones are insulating phases with integer band fillings $\nu_{e}$ marked in diagrams. In the phase diagrams, we keep the magnetic filling factor $\nu_{M}=1/2p$ fixed when tuning $\mu_{CF}$ to control the electron density by matching the magnetic field strength to the density ($B=2pn_e\phi_0$). All the insulating phases shown in the phase diagrams have a Chern number $\mathcal{C}=-1$.} 
\label{fig:PD}
\end{figure}

Figure~\ref{fig:PD} shows the phase diagram of CFs in the parameter plane of $V_{0}$--$\mu_{CF}$ for both the filling factors $1/2$ ($p=1$) and $1/4$ ($p=2$), where $\mu_{CF}$ is the CF chemical potential which determines the electron density and can be controlled by the external gates of the 2DEG device.  By tuning $\mu_{CF}$, we can make the band filling factor $\nu_{e}$ to be an integer, and obtain CF-QAHI when $V_0>V_c$.  We find the parameter regimes of insulating phase for $\nu_{e}$ up to $3$. All the insulating phases we find have a total Chern number $\mathcal{C}=-1$. We note that the system at $\nu_M=1/4$ ($p=2$) has the smaller $V_{c}$ than that at $\nu_M=1/2$ ($p=1$), suggesting that the system can be more easily transformed to a CF-QAHI at the stronger external magnetic field. The critical values $V_{c}$ for various magnetic filling factors at $\nu_{e}=1$ and their ratios to the Landau level spacing $\hbar\omega_0$ are listed in Table.~\ref{table:scales}. 

When in the insulating phase, the system will show the fractional quantum Hall effect. One could observe a quantum Hall plateau even though the system has an even-denominator magnetic filling factor. The relation between CFs' Chern number $\mathcal{C}$ and Hall conductance $\sigma^{e}_{xy}$ is
\begin{align}
\sigma^{e}_{xy}=-\frac{e^{2}}{h}\frac{\mathcal{C}}{2p\mathcal{C}+1}.
\label{CF-energybandequation}
\end{align} 
The unusual relation is due to the Faraday effect induced by the quantum vortices bound to the CFs~\cite{HLR,Fradkin1990}. Because all the insulating phases we find has a CF Chern number $\mathcal{C}=-1$, the system will be fractionally quantized at $-1/(2p-1)(e^{2}/h)$. For $\nu_{M}=1/2$ ($p=1$), $\sigma^{e}_{xy}=-e^{2}/h$, for $\nu_{M}=1/4$ ($p=2$), $\sigma^{e}_{xy}=-1/3(e^{2}/h)$. The experimental observation of an $1/3$ Hall plateau at $\nu_{M}=1/4$ will be a non-ambiguous indication for the forming of the CF-QAHI.

We further propose a many-body wave function for the CF-QAHI by generalizing the scheme described in Ref.~\onlinecite{Jains}.  For $\nu_{e}=1$, we propose,    
\begin{align}
\Psi=\hat{\mathcal{P}}_{MLLL}\Phi_{CF}\Phi^{2p}_{MLLL}
\label{CFstate}
\end{align}
where, $\Phi_{CF}$ is the mean-field wave function of CFs, and $\Phi_{MLLL}$ is the modified wave function of a filled lowest Landau level (LLL) in the presence of the periodic potential at the given electron density and under an reduced external magnetic field $\bm{B}/2p$, and $\hat{\mathcal{P}}_{MLLL}$ is a projection to the group of sub-bands originated from the lowest Landau level, denoted by $L=0$ in Fig~\ref{fig:MEB}. When the periodic potential is switched off, the wave function becomes the usual Rezayi-Read wave function for describing even-denominator fractional filling state~\cite{Rezayi1994}.  

We evaluate the energy stability of the CF state using the wave-function Eq.~(\ref{CFstate}). We employ the Metropolis Monte Carlo method to calculate kinetic energy and Coulomb energy~\cite{Ceperley1977}, and compare the total energy to a reference Hofstadter state of normal electrons at the same magnetic filling $\nu_{M}$ and band filling $\nu_{e}$. To simplify the calculation, we ignore the projection $\hat{\mathcal{P}}_{MLLL}$, although one expects the projected wave function would yield the lower expectation value of the energy~\cite{Jains}. The evaluation of the energy in the fully projected state will require developing a scheme in the modified LLL~\cite{Girvin1984}, which is not yet available and will be a topic for future research.   

Table~\ref{table:energy} shows the numerical results of the energy differences between the CF-state and the normal electron Hofstadter state.  One sees that while CF states have the higher kinetic energies, their Coulomb interaction energies can be significantly lowered. By comparing the total energy difference $\Delta E_{total}$, we conclude that the CF states can be stabilized when $r_{s}>r_{sc}$, where $r_{sc}$ ranges from $6.7$ to $17$ for the given set of parameters, depending on the filling factor $1/2p$.  We note that our calculation can only serve as a rough estimate of $r_{sc}$. The more accurate calculation taking full account of the band projection should further lower the total energies of CF-states, yielding smaller $r_{sc}$.

\begin{table}[tbp]
\caption{The kinetic energy difference $\Delta E_{K}$ and Coulomb interaction energy difference $\Delta V_{ee}$ between the CF state and the reference Holfstadter state, in the unit of $\epsilon_a$. The total energy difference $\Delta E_{total}= -\alpha(r_{s}-r_{sc})\epsilon_{a}$, with the parameters $\alpha$ and $r_{sc}$ listed in the table. In the calculation, we use the parameters $V_{0}=0.07\epsilon_{a}$, $\nu_{e}=1$.}
\begin{ruledtabular}

\begin{tabular}{|c||c|c|c|c|}
\hline
 $p$ & $\Delta V_{ee}$ & $\Delta E_{K}$ & $\alpha$ & $r_{sc}$  \tabularnewline
 \hline

 $p=1$ & $-0.007r_{s}$  & $0.047$ & $0.007$&$6.7 $  \tabularnewline
 \hline
 $p=2$ & $-0.009r_{s}$ & $0.071$ & $0.009$& $7.9$  \tabularnewline
 \hline
 $p=3$ & $-0.010 r_{s}$ & $0.110  $ & $0.010$ & $11$  \tabularnewline
 \hline
 $p=4$ & $-0.010r_{s}$ & $0.1276$ & $0.010$ & $13$ \tabularnewline
 \hline
 $p=5$ & $ -0.011r_{s}$ & $0.1826$ & $0.011$ & $16$ \tabularnewline

 \end{tabular}
 \end{ruledtabular}
 \label{table:energy}
 \end{table}

Finally, we discuss the experimental realization of our proposal.  For the given magnetic filling factor $\nu_{M}=1/2p$ and integer band filling factor $\nu_{e}$, once we choose a lattice constant $a$ of the hexagonal periodic potential, both the electron density $n_{e}=2\nu_{e}/(3\sqrt{3}a^{2})$ and the external magnetic field strength ${B}=4p\phi_{0}\nu_{e}/(3\sqrt{3}a^{2})$ are fixed. The constraints on the value of $a$ are both the technical feasibility of nano-fabrications and the requirement that the electron density should be low enough such that $r_{s}>r_{sc}$. With these in mind, for a 2DEG in GaAs-AlGaAs heterostructure, we can choose $a=100\mathrm{nm}$, $n_{e}=0.38\times10^{10}/\mathrm{cm}^{2}$,  and $V_{0}=0.19\mathrm{meV}$. With $r_{s}=7.6$, the CF-QAHI can be stabilized for $p=1$ ($B=0.157\mathrm{T}$). The extremely low electron density could be challenging but not unreachable~\cite{Kane1993}. A more realistic set of parameters could be obtained in the hole system~\cite{Tsui1992}, in which the effective mass is much larger, yielding much larger $r_s$ than its electron counterpart at the same carrier density. In this case, we can choose $a=30\mathrm{nm}$, $n_{h}=0.427\times10^{11}/\mathrm{cm}^{2}$ and $V_{0}=0.40\mathrm{meV}$. We can have a hole system with $r_{s}=12.69$ ($m_{b}=0.3m_{e}$) that is sufficient to stabilize CF-QAHI phases up to $p=3$.     

In summary, we propose CF-QAHI, a new class of QAHI exhibiting the fractional topological order.  We show that a weak hexagonal periodic potential could transform a 2DEG system at even-denominator magnetic filling to CF-QAHI. The experimental realization of the state, while challenging, is a natural extension of recent efforts in constructing artificial Dirac-fermion  
system in 2DEG~\cite{Polini2013} and could be a more practical scheme in finding a QAHI with fractional quantum Hall effect.
   
We gratefully acknowledge the useful discussions with Xi Lin and Chi Zhang, and the supports from the National Basic Research Program of China (973 Program) Grant No. 2012CB921304, and from National Science Foundation of China (NSFC) Grant No. 11325416.


\newpage

\vskip 15 cm
\begin{widetext}
\setcounter{figure}{0}
\setcounter{equation}{0}
\setcounter{section}{0}

\renewcommand\thefigure{S\arabic{figure}}
\renewcommand\theequation{S\arabic{equation}}

\section{Supplementary information for {}`Quantum `Anomalous Hall Insulator of Composite Fermions''}

\vskip 0.5 cm
In this supplemental material, we provide the details of the calculation in the letter, including the calculation of Hofstadter spectrum structure in Figure~(2), the construction of a many-body wave function in Eq.~(5), the total energy calculation of the system in Table~II.  We also include a section on Chern-Simons current-density functional theory, which could provide a theoretical basis for the CF-mean field theory employed in the main text.

\subsection{The calculation of Hofstadter spectrum structure}
In this section, we show the details about the calculation of Hofstadter spectra in Fig.~2(a). Consider a non-interacting 2DEG system on a weak periodic background with a magnetic filling factor $\nu_{M}=l/h$ (with $l$ and $h$ being integers) and the band filling factor $\nu_{e}=1$.  The Hamiltonian is,
\begin{equation}
H=\frac{1}{2m_{b}}(\bm{p}-\frac{e}{c}\bm{A})^{2}+U(\bm{r})
\end{equation}
where $U(\bm{r})$ is defined in Eq.~(1), and $\bm{A}=1/2(-By,Bx)$ is the corresponding vector potential of the external magnetic field $B$. 

One can define a magnetic translation operators $t_{m}(\bm{R})=\exp(i \boldsymbol{\kappa}\cdot \bm{R}/\hbar)$, with
\begin{equation}
\boldsymbol{\kappa}=(\bm{p}-\frac{e}{c}\bm{A})-\hbar \frac{\hat{z}\times \bm{r}}{l^{2}_{m}}.
\end{equation}
and the crystal lattice vector $\bm{R}$ spanned by the unit lattice vectors $\bm{a}_{1}=(\sqrt{3}/2,3/2)a$ and $\bm{a}_{2}=(\sqrt{3},0)a$, where $l_{m}$ is the corresponding magnetic length. The magnetic translation operators do not commutate with each other or with Hamiltonian $H$. As a result, one needs to enlarge the unit cell to a supperlattice enclosed by $2l\bm{a}_{1}$ and $\bm{a}_{2}$, denoted as a magnetic unit cell. The choice of $2$ is due to the shape of the unit cell (See Ref. [28] in the main text). In the corresponding magnetic Brillouin zone (MBZ), one can define a set of Bloch basis wave functions with the quasi-momentum $\bm{\tilde{q}}$:
\begin{equation}
B^{n_{l}}_{j,\tilde{\bm{q}}}(x,y)=\sum_{m}M_{j,m}(\tilde{q}_{y})\bar{\psi}_{n_{l},\tilde{q}_{x}+(2mh+j)b_{2x}}(x,y), \quad j=0,1,..,2h-1,
\end{equation}
where
\begin{equation}
\bar{\psi}_{n_{l},q_{x}}(x,y)=A_{n_{l}}H_{n_{l}}\left(y-q_{x}l^{2}_{m}\right)\exp \left(-\frac{1}{2l^{2}_{m}}\left(y-q_{x}l^{2}_{m}\right)^{2}\right)\exp \left(-i \frac{xy}{2 l_{m}^{2}}\right)\exp \left(i q_{x} x \right),
\end{equation}
with $n_{l}$ being the Landau index, $H_{n_{l}}(x)$ being Hermite polynomials and $A_{n_{l}}$ being a normalized coefficient, and
\begin{equation}
M_{j,m}(\tilde{q}_{y})=\exp \left(i\tilde{q}_{y}a_{1y}(2ml+j\frac{l}{h})\right).
\end{equation}

The secular equation based on the basis functions $\{B^{n_{l}}_{j,\tilde{\bm{q}}}(x,y)\}$ is:\begin{equation}
\sum_{n'_{l},j'}\left [\left (n_{l}+\frac{1}{2}-E\right)\delta_{n_{l},n'_{l}}\delta_{j,j'}+V^{n_{l},n'_{l}}_{j,j'}(\tilde{\bm{q}}) \right] C_{n'_{l},j'}(\tilde{\bm{q}})=0,
\label{MEB}
\end{equation}
where $V^{n'_{l},n_{l}}_{j',j}(\tilde{\bm{q}})\equiv\langle B^{n'_{l}}_{j',\tilde{\bm{q}}}|U(\bm{r})|B^{n_{l}}_{j,\tilde{\bm{q}}} \rangle$ is the periodic potential matrix element. To calculate the matrix elements, one can utilize:
\begin{equation}
\langle B^{n'_{l}}_{j',\tilde{\bm{q}}}|\exp(i\bm K \cdot \bm r)|B^{n_{l}}_{j,\tilde{\bm{q}}} \rangle =F_{n'_{l},n_{l}}(\bm{K})\exp \left(iK_{y}(\tilde{q}_{x}+j b_{2x})l^{2}_{m}\right)\exp \left(-i \tilde{q}_{y}a_{1y}\frac{l}{h}\frac{K_{x}}{b_{2x}} \right)\,\, \mathrm{when}\,\, j=j'- n_2 + 2 h N ,
\end{equation} 
where $\bm K = n_1 \bm b_1 + n_2 \bm b_2$ is a reciprocal lattice vector, $N \in Z$, and, 
\begin{equation}
F_{n'_{l},n_{l}}(\bm{K})=\sqrt{\frac{\nu_{1}!}{\nu_{2}!}}\left(\frac{(sgn(n_{l}-n'_{l})K_{x}+iK_{y})l_{m}}{\sqrt{2}}\right)^{\xi}L^{\xi}_{\nu_{1}}\left(\frac{K^2l^{2}_{m}}{2}\right)\exp \left(-\frac{K^{2}l^{2}_{m}}{4}+\frac{iK_{x}K_{y}l^{2}_{m}}{2}\right),
\end{equation}
with the generalized Laguerre polynomial $L^{\xi}_{\nu_{1}}(x)$, and $\xi=|n_{l}-n'_{l}|$, $\nu_{1}=\min(n_{l},n'_{l})$, $\nu_{2}=\max(n_{l},n'_{l})$, sgn(x) being sign function. After the diagonalization of Eq.~(\ref{MEB}), we can determine the Hofstadter wave functions:
\begin{equation}
 \phi_{s,\bm{q}}(x,y)=\sum_{n'_{l},j'} C^{s}_{n'_{l},j'}(\tilde{\bm{q}})B^{n'_{l}}_{j',\tilde{\bm{q}}}(x,y),
 \end{equation}
with $s$ being the Hofstadter energy band index.

\subsection{The construction of many-body wave functions}

The construction of many-body wave function for the CF-QAHI in Eq.~(5) involves two part contributions:  the CF mean field wave function $\Phi_{CF}$ and the modified lowest Landau level (MLLL) wave function $\Phi_{MLLL}$. 

To get $\Phi_{CF}$, we solve Eq.~(2), and obtain a set of CF Bloch wave functions $\psi^{CF}_{n,\bm{k}}$, When $N$ CF-particles are fully occupied in a lowest CF energy band $n=1$, we can select the set of the occupied CF Bloch wave functions $\{ \psi^{CF}_{n,\bm{k}}(\bm{r})| n=1;\bm{k} \in BZ\}$ to build the $N$ CF-particle Slater determinant wave function
\begin{align}
\Phi_{CF}& =\left|\begin{array}{cccc}
\psi_{n=1,\bm{k}_{1}}^{CF}(\bm{r}_{1}) & \psi_{n=1,\bm{k}_{1}}^{CF}(\bm{r}_{2}) & \cdots & \psi_{n=1,\bm{k}_{1}}^{CF}(\bm{r}_{N})\\
\psi_{n=1,\bm{k}_{2}}^{CF}(\bm{r}_{1}) & \psi_{n=1,\bm{k}_{2}}^{CF}(\bm{r}_{2}) & \cdots & \psi_{n=1,\bm{k}_{2}}^{CF}(\bm{r}_{N})\\
\vdots & : & \ddots & \vdots\\
\psi_{n=1,\bm{k}_{N}}^{CF}(\bm{r}_{1}) & \psi_{n=1,\bm{k}_{N}}^{CF}(\bm{r}_{2}) & \cdots & \psi_{n=1,\bm{k}_{N}}^{CF}(\bm{r}_{N})
\end{array}\right|\equiv\det{[\psi^{CF}_{m}(\bm{r}_{n})]},
\end{align}
where $\bm k$ is defined in a $76\times 76$ mesh of BZ in our calculation.  

For $\Phi_{MLLL}$, we select the lowest $N$ Hofstadter wave functions $\{\phi_{s,\tilde{\bm{q}}}(x,y)|s=1,2; \tilde{\bm{q}}\in MBZ \}$  at $\nu_{M}=1$ to build $N$ particle many-body function 
\begin{align}
\Phi_{MLLL} & =\left|\begin{array}{ccccc}
\phi_{s=1,\tilde{\bm{q}}_{1}}(\bm{r}_{1}) & \cdots & \phi_{s=1,\tilde{\bm{q}}_{1}}(\bm{r}_{j}) & \cdots & \phi_{s=1,\tilde{\bm{q}}_{1}}(\bm{r}_{N})\\
\vdots & \vdots & \vdots  & \cdots & \vdots\\
\phi_{s=1,\tilde{\bm{q}}_{N/2}}(\bm{r}_{1}) & \cdots & \phi_{s=1,\tilde{\bm{q}}_{N/2}}(\bm{r}_{j}) & \cdots & \phi_{s=1,\tilde{\bm{q}}_{N/2}}(\bm{r}_{N})\\
\phi_{s=2,\tilde{\bm{q}}_{N/2+1}}(\bm{r}_{1}) & \cdots & \phi_{s=2,\tilde{\bm{q}}_{N/2+1}}(\bm{r}_{j}) & \cdots & \phi_{s=2,\tilde{\bm{q}}_{N/2+1}}(\bm{r}_{N})\\
\vdots & \vdots & \vdots & \vdots & \vdots\\
\phi_{s=2,\tilde{\bm{q}}_{N}}(\bm{r}_{1}) & \cdots & \phi_{s=2,\tilde{\bm{q}}_{N}}(\bm{r}_{j}) & \cdots & \phi_{s=2,\tilde{\bm{q}}_{N}}(\bm{r}_{N})
\end{array}\right|\equiv\det{[\phi_{m}(\bm{r}_{n})]}.
\end{align}
We use the many-body function Eq.~(5) without MLLL projection to calculate the system energy.  

\subsection{The Monte-Carlo calculation of energy}
In this section, we show the Monte-Carlo evaluation of the ground state energy. The total energy has two parts of contributions, one is from the kinetic energy, defined here as all the contributions from the single-body part of Hamiltonian; the other is from the Coulomb energy, which could be calculated straightforwardly.
  
 The average total kinetic energy for $N$ particle system is expressed as
\begin{align}
t & =\frac{1}{AN}\sum^{N}_{j=1}\int\prod_{i}d\bm{r}_{i}\Psi^{\ast}(\bm{r}_{1},...,\bm{r}_{N})\left (\frac{1}{2m_{b}}(\bm{p}_{j}-\frac{e}{c}\bm{A}_{j})^{2}+U(\bm{r}_{j})\right)\Psi(\bm{r}_{1},..,\bm{r}_{N}),\label{Kinetic}
\end{align}
in which $\Psi(\bm{r}_{1},..,\bm{r}_{N})$ is many-body wave function for CF-QAHI in Eq.~(5), and $A=\int\prod_{i}d\bm{r}_{i}|\Psi(\bm{r}_{1},...,\bm{r}_{N})|^{2}$ is the normalized coefficient. Using the Monte Carlo approach, we sample particle coordinators with the probability $|\Psi(\bm{r}_{1},...,\bm{r}_{N})|^{2}/A$. We have: 
\begin{align}
t & =\frac{1}{N_{s}}\sum^{N_{s}}_{s=1}\frac{1}{N}\sum^{N}_{j=1} \left \{ \frac{\hbar^{2}}{2m_{b}}\left |\nabla_{j}\ln\Psi(\bm{r}^{s}_{1}, \bm{r}^{s}_{2},..., \bm{r}^{s}_{N})-i\frac{e}{\hbar c}\bm{A}(\bm{r}^{s}_{j})\right |^{2}+U(\bm{r}^{s}_{j})\right\},\label{Kinetic2}
\end{align}
in which $N_{s}$ is the total number of samplings, and $\{{\bm{r}^{s}_{1},..,\bm{r}^{s}_{N}}\}$ is the spatial coordinator set of the $s$-th sampling. 

For the CF many-body wave function Eq.~(5),  we have: 
\begin{multline}
t  =\frac{1}{N_{s}}\sum_{s=1}^{N_{s}}\frac{1}{N}\sum_{j=1}^{N}\left[\frac{\hbar^{2}}{2m_{b}}\left(\sum_{i=1,2}\left|\det[(\psi_{m}^{CF}(\bm{r}_{n}^{s}))_{j_{1}j_{2}}^{-1}]\nabla_{j}\det[(\psi_{m}^{CF}(\bm{r}_{n}^{s}))_{j_{2}j_{3}}]\right.\right.\right. \notag \\
  \left.\left.\left.+2p\det[(\phi_{m}(\bm{r}_{n}^{s}))_{j_{1}j_{2}}^{-1}]\nabla_{j}\det[(\phi_{m}(\bm{r}_{n}^{s}))_{j_{2}j_{3}}]-\frac{ie}{\hbar c}\bm A(\bm{r}_{j}^{s})\right|{}^{2}\right)+U(\bm{r}_{j}^{s})\right],\label{Kinetic3}
\end{multline}
where $\nabla_{j}$ represents the gradient of the $j$-th particle, the $N \times N$ matrix $(\psi_{m}^{CF}(\bm{r}_{n}^{s}))_{j_{1}j_{2}}$ has the element $\psi^{CF}_{j_{1}}(\bm{r}^{s}_{j_{2}})$. The matrix elements of $(\phi_{m}(\bm{r}^{s}_{n}))_{j_{2}j_{3}}$ consist of the Hofstadter wave function $\phi_{m}(\bm{r}_{n})$ at $\nu_{M}=1$. The values of determinants can be efficiently updated for using the propagative method presented in Ref. [32] of the main text.  
  
To calculate the Coulomb interaction energy,  we employ the Ewald summation formula for 2D system~\cite{Grzybowski2000},
\begin{equation}
V_{ee}=\frac{1}{2}\sum_{i,j}q_{i}q_{j}\psi(\bm{r}_{i}-\bm{r}_{j})-\frac{\alpha}{\sqrt{\pi}}\sum_{i=1}^{N}q_{i}^{2},\label{Coulomb}
\end{equation}
where $q_{i}$ is the charge of the $i$-th particle in unit of $|e|$, the last term is an interacting energy for background, and the first term is the electrostatic potential 
\begin{align}
\psi(\bm{r}_{i}-\bm{r}_{j}) & \equiv \sum_{\bm{n}_{\rho}}'\frac{erfc[\alpha((\bm{r}_{i}-\bm{r}_{j})+\bm{n}_{q})]}{|\alpha((\bm{r}_{i}-\bm{r}_{j})+\bm{n}_{q})|}+\frac{2\pi}{S}\sum_{\bm{G}\neq0}\frac{\exp(i\bm{G}\cdot(\bm{r}_{i}-\bm{r}_{j}))}{G}erfc(\frac{G}{2\alpha}).
\end{align}
with $erfc(x)$ being a complementary error function, and $\alpha$ being Ewald convergence parameter, and $\bm{n}_{q}$ ($\bm{G}$) being the direct (reciprocal) lattice of images. In our case, the positive charges form a uniform background (jellium model)~\cite{Osychenko2011}, the above formula is modified to:  
\begin{equation}
 V_{ee} =\frac{1}{2}\sum_{ij}^{N}\psi(\bm{r}_{ij})-N^{2}\frac{\sqrt{\pi}}{\alpha S}-\frac{\alpha}{\sqrt{\pi}}N,
\end{equation} 
where  $N$ is the total number of electrons, and $S$ is the total area of the system.

\subsection{Chern-Simons current-density functional theory (CS-CDFT)}

In this section, we sketch a Chern-Simons current-density functional theory (CS-CDFT), which could provide a justification to the mean field theory employed in the main text. In general, we are treating systems described by the Hamiltonian:
\begin{align}
\hat{H}_{e}=\sum_{j}\left[\frac{1}{2m_{b}}\left(-i\hbar \bm \nabla_j-\frac{e}{c}\bm{A}(\bm{r}_{j})\right)^{2}+U(\bm{r}_{j})\right]+\frac{1}{2}{\sum_{i,j}}^\prime V_{ee}(\bm{r}_{i}-\bm{r}_{j}),
\end{align}
where $V_{ee}(\bm{r}-\bm{r}')$ is electron-electron interaction. By making a singular Chern-Simons gauge transformation that transforms the many-body wave function by:
\begin{align}
\Psi(\bm{r}_{1},..\bm{r}_{N})\rightarrow\exp(-2pi\sum_{i<j}\arg(\bm{r}_{i}-\bm{r}_{j}))\Psi(\bm{r}_{1},..,\bm{r}_{N}),
\end{align}
where $\arg(\bm{x})$ is the angle made between the vector $\bm{x}$ and the positive real axis, the Hamiltonian is transformed to,
\begin{align}
\hat{H}_{CF}=\sum_{j}\left[\frac{1}{2m_{b}}\left(\hat{\bm{p}}_{j}-\frac{e}{c}\bm{A}_0(\bm{r}_{j})-\frac{e}{c}\bm{A}^\prime(\bm{r}_{j})-\frac{e}{c}\frac{p\phi_0}{\pi}\sum_{i\ne j} \frac{\hat{z}\times(\bm r_j - \bm r_i)}{|\bm r_j - \bm r_i|^2}\right)^{2}+U(\bm{r}_{j})\right]+\frac{1}{2}{\sum_{i,j}}^\prime V_{ee}(\bm{r}_{i}-\bm{r}_{j}), \label{TransformedH}
\end{align}
where we decompose the vector potential $\bm A(\bm r)$ into two parts such that
\begin{align}
\bm \nabla \times \bm A_0(\bm r) = 2p\phi_0 \bar{n}_e,
\end{align}
because we are only interested in the regime near the filling factor $\nu = 1/2p$, and the decomposition will make the resulting $\bm A^\prime(\bm r)$ small.  We should find the ground states of the interacting systems described by (\ref{TransformedH}), for different vector potentials $\bm{A}^\prime(\bm{r})$ and scalar potentials $U(\bm{r})$.  We will show that it is possible to set up a non-interacting Kohn-Sham equation, which will give rise to a set of single particle wave functions that could be interpreted as the composite fermion wave functions.  Moreover, the mapping is exact in principles. 

We follow the general scheme of derivation presented in Ref.~\cite{Vignale1987}.  First of all, one can establish a generalized Kohn-Hohnberg theorem, which states that the non-degenerate ground state wave function $\Psi$ as well as $\bm{A}^\prime(\bm{r})$ and $U(\bm{r})$ are uniquely determined by the distributions of the density $n(\bm r)$ and ``paramagnetic'' current density $\bm j_p(\bm r)$, apart from a trivial additive constant in the scalar potential. Here, the ``paramagnetic'' current density is defined as:
\begin{equation}
\bm j_p(\bm r_1) = \int \mathrm{d}\bm r_2 \dots \mathrm{d}\bm r_N \left[
-\frac{i\hbar}{2m_b}\left(\Psi^* (\bm\nabla_1 \Psi) - (\bm\nabla_1 \Psi^*)  \Psi\right) - 
\frac{e}{c}\left(A_0(\bm r_1) +\frac{p\phi_0}{m_b\pi}\sum_{i\ne 1} \frac{\hat{z}\times(\bm r_1 - \bm r_i)}{|\bm r_1 - \bm r_i|^2}\right) |\Psi|^2 \right]
\end{equation}

As a result, the ground state energy must be a functional of $n(\bm r)$ and $\bm j_p(\bm r)$:
\begin{align}
E_{V,\bm{A}^\prime}[n,\bm{j}_{p}]=F[n,\bm{j}_{p}]-\frac{e}{c}\int d\bm{r}\bm{j}_{p}(\bm{r})\cdot\bm{A}^\prime(\bm{r})+\int d\bm{r}n(\bm{r})\left[U(\bm{r}) + \frac{e^{2}}{2m_{b}c^{2}}\bm{A}^{\prime2}(\bm{r})\right].
\end{align}

Furthermore, one can decompose the functional $F[n,\bm{j}_{p}]$ as:
\begin{equation}
F[n,\bm{j}_{p}] = T_s[n, \bm{j}_{p}] + \frac{1}{2}\int \int \mathrm{d} \bm r \mathrm{d}\bm r^\prime n(\bm r)V_{ee}(\bm r - \bm r^\prime) n(\bm r^\prime) + E_{xc}[n, \bm{j}_{p}], \label{F}
\end{equation}
with $T_s[n, \bm{j}_{p}]$ being a functional of \emph{non-interacting} systems, defined as,
\begin{equation}
T_s [n, \bm{j}_{p}] = \left\langle\Psi_0[n,\bm j_p]\left| \sum_i\frac{\left[-i\hbar\bm\nabla_i - \frac{e}{c}\left(\bm A_0(\bm r_i)+\bm a_{CS}(\bm r_i)\right)\right]^2}{2m_b}\right|\Psi_0[n,\bm j_p]\right\rangle,
\end{equation}
where $\Psi_0$ is the ground state wave function of a $N$-particle \emph{non-interacting} system with the given distributions of $n(\bm r)$ and $\bm j_p(\bm r)$,  and $\bm a_{CS}(\bm r)$ is determined by:
\begin{equation}
\nabla \times \bm{a}_{CS}(\bm{r})=-2p\phi_{0}n(\bm{r}).
\end{equation}

The ground state wave-function $\Psi_0$ of the non-interacting reference system can be constructed from $N$ lowest-lying solutions of a single-body Schr\"odingier equation with effective vector and scalar potentials $\bm A_{eff}(\bm r)$, $U_{eff}(\bm r)$:
\begin{equation}
\left[\frac{\left(-i\hbar\bm\nabla - \frac{e}{c}\bm A_{eff}(\bm r)\right)^2}{2m_b} + U_{eff}(\bm r)\right]\psi_i(\bm r) = \epsilon_i \psi_i(\bm r), \label{Sch}
\end{equation}
where $\bm A_{eff}(\bm r)$ and $U_{eff}(\bm r)$ are functionals of $n(\bm r)$ and $\bm j_p(\bm r)$, and are chosen such that
\begin{eqnarray}
n(\bm r) &=& \sum_{i=1}^N |\psi_i(\bm r) |^2 , \\
\bm{j}_{p}(\bm{r}) &=& -\sum_{i=1}^{N}\frac{i\hbar}{2m_b}(\psi_{i}^{\ast}\bm\nabla\psi_{i}-c.c) - \frac{e}{m_b c} n(\bm r) \left(\bm A_0(\bm r) + \bm a_{CS}(\bm r)\right).
\end{eqnarray} 

The functional $T_s[n,\bm j_p]$ can then be re-expressed as,
\begin{multline}
T_s[n,\bm j_p] = \sum_{i=1}^N \epsilon_i - \frac{e}{c} \int \mathrm{d}\bm r \left[\bm A_0(\bm r)  +\bm a_{CS}(\bm r) - \bm A_{eff}(\bm r)\right]\cdot \left[\bm j_p(\bm r) +\frac{e}{m_b c}n(\bm r) \left(A_0(\bm r) + a_{CS}(\bm r)\right)\right] \\
+ \int \mathrm{d}\bm r  n(\bm r)\left[ \frac{e^2}{2m_b c^2} \left(\left(\bm A_0(\bm r)+\bm a_{CS}(\bm r)\right)^2 - \bm A_{eff}^2(\bm r) \right) - U_{eff}(\bm r) \right], \label{T_s}
\end{multline}
where $\epsilon_i$, $\bm a_{CS}(\bm r)$, $\bm A_{eff}(\bm r)$, $U_{eff}(\bm r)$ are all functionals of $n$ and $\bm j_p$.

Substituting Eq.~(\ref{T_s}) into Eq.~(\ref{F}), and minimizing the total energy functional against $n_p$ and $\bm j_p$, one can determine the effective scalar and vector potentials:
\begin{align}
\bm A_{eff}(\bm r) & =  \bm a_{CS}(\bm r) +\bm A_0(\bm r)  +\bm A^\prime(\bm r) + \bm  A_{xc}(\bm r)  ,\label{Aeff}\\
 U_{eff}(\bm r)  & =  U(\bm r)  +U_{xc}(\bm r) + 2p\phi_0 m_z(\bm r)  +\int \mathrm{d}\bm r^\prime V_{ee}(\bm r-\bm r)n(\bm r) \nonumber \\
& + \frac{e^2}{2m_b c^2} \left[\bm A^{\prime2}(\bm r) - \left(\bm A^\prime(\bm r) + \bm  A_{xc}(\bm r)\right)^2\right] , \label{Ueff}
\end{align}
where $m_z(\bm r)$ is the orbital magnetization density of the non-interacting system defined in Eq.~(\ref{Sch}), $U_{xc}(\bm r) = \delta E_{xc}/\delta n(\bm r)$, $\bm A_{xc}(\bm r) = - (c/e) \delta E_{xc}/\delta \bm j_p(\bm r)$. 

Equations (\ref{Sch}), (\ref{Aeff}), (\ref{Ueff}) define a self-consistent non-interacting problem, which gives rise to the same density and current density as the true man-body ground state.  When we ignore the corrections due to the exchange-correlation effects, the equation actually has the same form as the CF-mean field employed in the main text.  The presence of the exchange-correlation effects will in general renormalize the effective scalar and vector potentials, as indicated by Eqs.~ (\ref{Aeff}), (\ref{Ueff}).  We note that the mapping of a complex many body problem defined in Eq.~(\ref{TransformedH}) to the non-interacting problem is exact, provided that one has the exact knowledges of the functionals.  In practice, one has to employ approximation (e.g., the local density approximation).  In this case, one expects that the solution should provide a sensible description to the CF-ground state in the weak potential limit.

\end{widetext}


\begin{thebibliography}{99}

\bibitem{Haldane-model} F.D.M. Haldane, Phys. Rev. Lett \textbf{61}, 2015 (1988)

\bibitem{Onoda2003} M. Onoda, N. Nagaosa, Phys. Rev. Lett. \textbf{90}, 206601 (2003) 

\bibitem{Liu2008} C.-X. Liu, X.L. Qi, X. Dai, Z. Fang, and S.C. Zhang, Phys. Rev. Lett. \textbf{101}, 146802 (2008).

\bibitem{Yu2010} R. Yu, W. Zhang, H.-J. Zhang, S.-C. Zhang, X. Dai, and Z. Fang, Science 329, 61 (2010).

\bibitem{qiao2010} Z. Qiao, S. A. Yang, W. Feng, W.-K. Tse, J. Ding, Y. Yao, J. Wang, and Q. Niu, Phys. Rev. B \textbf{82}, 161414(R) (2010).

\bibitem{qiao2012} Z. Qiao, H. Jiang, X. Li, Y. Yao, and Q. Niu, Phys. Rev. B \textbf{85}, 115439 (2012).

\bibitem{zhang2012} H. Zhang, C. Lazo, S. Blugel, S. Heinze, and Y. Mokrousov, Phys. Rev. Lett. \textbf{108}, 056802 (2012).

\bibitem{chang2013} C. -Z. Chang {\it et al.}, Science \textbf{340}, 6129 (2013)

\bibitem{Sheng2011} D. N. Sheng, Zheng-Cheng Gu, Kai Sun and L. Sheng, Nature commun. \textbf{2}, 389 (2011) 

\bibitem{Tang2011} E. Tang, J.-W. Mei, X.-G. Wen, Phys. Rev. Lett. \textbf{106}, 236802 (2011)

\bibitem{Sun2011} K. Sun, Z.-C. Gu, H. Katsura, S. Das Sarma, Phys. Rev. Lett. \textbf{106}, 236803 (2011)

\bibitem{Neupert2011} T. Neupert, L. Santos, C. Chamon, C. Mudry, Phys. Rev. Lett. \textbf{106}, 236804 (2011) 


\bibitem{Bergholtz2013} Emil J. Bergholtz and Zhao Liu, Int. J. Mod. Phys. B 27, 1330017 (2013)







\bibitem{Jains} J. K. Jain, 2007, {\it Composite Fermions} (Cambridge University Press, 2007)

\bibitem{Heinonen} O. Heinonen, ed., {\it Composite Fermions} (World Scientific, 1998).

\bibitem{Wen2004} Xiao-Gang Wen, {\it Quantum Field Theory of Many-Body Systems} (Oxford University Press, 2004) 

\bibitem{HLR} B. I. Halperin, P. A. Lee, and N. Read, Phys. Rev. B \textbf{47}, 7312 (1993). 

\bibitem{Kalmeyer1992} V. Kalmeyer and S. C. Zhang, Phys. Rev. B \textbf{46}, 9889 (1992). 

\bibitem{Willett1993} R.L. Willett, R.R. Ruel, K.W. West, and L.N. Pfeiffer, Phys.Rev. Lett. \textbf{71}, 3846 (1993)

\bibitem{Kang1993a} W. Kang, H. L. Stormer, L. N. Pfeiffer, K. W. Baldwin, and K. W. West, Phys. Rev. Lett. \textbf{71}, 3850 (1993)

\bibitem{Goldman1994} V.J. Goldman, B. Su, and J.K. Jain, Phys. Rev. Lett. \textbf{72}, 2065 (1994).

\bibitem{Smet1996} J. H. Smet, D. Weiss, R. H. Blick, G. Lutjering, K. von Klitzing, R. Fleischmann, R. Ketzmerick, T. Geisel, and G. Weimann, Phys. Rev. Lett. \textbf{77}, 2272 (1996) 

\bibitem{Smet1999} J.H. Smet, S. Jobst, K.von Klitzing, {\it et al.}, Phys. Rev. Lett. \textbf{83}, 2620 (1999)

\bibitem{Oppen1998} F. von Oppen, Ady Stern, and Bertrand I. Halperin, Phys.Rev. Lett. \textbf{80}, 4494 (1998)

\bibitem{Mirlin1998} A. D. Mirlin, D. G. Polyakov and P. Wolfle, Phys .Rev. Lett. \textbf{80}, 2429 (1998)







\bibitem{TKNN} D. J. Thouless, M. Kohmoto, M. P. Nightingale, and M. den Nijs, Phys. Rev. Lett. \textbf{49}, 405 (1982).

\bibitem{Hofstadter1976} D. R. Hofstadter, Phys. Rev. B \textbf{14}, 2239 (1976).

\bibitem{MacDonald1984} A. H. MacDonald, Phys. Rev. B \textbf{29}, 3057 (1984)


\bibitem{Xiao2010} Di Xiao, Ming-Che Chang and Qian NIu, Rev. Mod. Phys. \textbf{82},1959 (2010)

\bibitem{Fradkin1990} Ana Lopez and E. Fradkin, Phys. Rev. B \textbf{44}, 5246 (1991)

\bibitem{Rezayi1994} E. Rezayi and N. Read, Phys. Rev. Lett. 72, 900 (1994).

\bibitem{Ceperley1977} D. Ceperley, G. V. Chester, and M. H. Kalos, Phys. Rev. B \textbf{16}, 3081 (1977) 

\bibitem{Girvin1984} S. M. Girvin and Terrence Jach, Phys. Rev. B. \textbf{29}, 5617 (1984)

\bibitem{Kane1993} B. E. Kane, L. N. Pfeiffer, K. W. West, and C. K. Harnett, Appl. Rev. Lett. \textbf{63}, 2132 (1993)

\bibitem{Tsui1992} M. B. Santos, Y. W. Suen, M. Shayegan, Y. P. Li, L. W. Engel, and D. C. Tsui, Phys. Rev. Lett. \textbf{68}, 1188 (1992) 


\bibitem{Polini2013} Marco Polini, Francisco Guinea, Maciej Lewenstein, Hari C. Manoharan and Vittorio Pellegrini, Nature Nanotech., \textbf{4} 625 (2013)






%


%

\end{thebibliography}

\begin{thebibliography}{References}

\bibitem{Grzybowski2000} A. Grzybowski, {\it et al.}, Phys. Rev. B \textbf{61}, 6706(2000).

\bibitem{Osychenko2011} O. N. Osychenko, G. E. Astrakharchik, and J. Boronat, arxiv:1107.5435v1.

\bibitem{Vignale1987} G. Vignale and M. Rasolt, Phys. Rev. Lett. \textbf{59}, 2360 (1987); Phys. Rev. B \textbf{37}, 10685 (1988).

\end{thebibliography}
\end{document}